\documentclass[prb,twocolumn,superscriptaddress,showpacs]{revtex4}
\usepackage{graphics}
\usepackage{graphicx}
\usepackage{amssymb}
\usepackage{amsmath}

\begin{document}

\title{Spin-Guide: A New Source of High Spin-Polarized Current}

\author{R.N.Gurzhi}
\affiliation{B.Verkin Institute for Low Temperature Physics \&
Engineering of the National Academy of Science of the Ukraine, 47
Lenin Ave, Kharkov, 61103, Ukraine}
\author{A.N.Kalinenko}
\affiliation{B.Verkin Institute for Low Temperature Physics \&
Engineering of the National Academy of Science of the Ukraine, 47
Lenin Ave, Kharkov, 61103, Ukraine}
\author{A.I.Kopeliovich}
\affiliation{B.Verkin Institute for Low Temperature Physics \&
Engineering of the National Academy of Science of the Ukraine, 47
Lenin Ave, Kharkov, 61103, Ukraine}
\author{A.V.Yanovsky}
\affiliation{B.Verkin Institute for Low Temperature Physics \&
Engineering of the National Academy of Science of the Ukraine, 47
Lenin Ave, Kharkov, 61103, Ukraine}
\author{E.N.Bogachek}
\affiliation{School of Physics, Georgia Institute of Technology,
Atlanta, GA 30332-0430, USA}
\author{Uzi Landman}
\affiliation{School of Physics, Georgia Institute of Technology,
Atlanta, GA 30332-0430, USA}

\begin{abstract}
We propose a ``spin-guide'' source for generation of electric
currents with a high degree of spin polarization, allowing
long-distance transmission of the spin-polarization. In the
spin-guide scheme proposed here, a non-magnetic conducting channel
is wrapped by a magnetic shell which preferentially transmits
electrons with a particular spin polarization. It is shown that
this method is significantly more effective then the
spin-filter-like scheme where the current flows perpendicular to
the interface between a ferromagnetic metal to a non-magnetic
conducting material. Under certain conditions a spin-guide may
generate an almost perfectly spin-polarized current, even when the
magnetic material used is not fully polarized. The spin-guide is
predicted to allow the transport of spin polarization over long
distances which may exceed significantly the spin-flip length in
the channel. In addition, it readily permits detection and control
of the spin-polarization of the current. The spin-guide may be
employed for spin-flow manipulations in semiconductors used in
spintronic devices.
\end{abstract}

\date{January 8, 2003}
\maketitle

\section{INTRODUCTION}

Recently there has been a growing interest in ``spintronic''
devices \cite{pr,ds}, where the spin degree of freedom is utilized
for data manipulations, rather then just the electronic charge as
in customary devices. This is due to the obvious advantages of
integrating a magnetic data storage device with an electronic
readout, as well as due to the promising prospects for
applications of spin-polarized currents in quantum computing. The
main technical requirements for the development of spintronic
devices, pertain to: (i) high efficiency spin injection into a
semiconductor, and (ii) long-distance propagation of the spin
signal. Currently, some of the major issues concerning the
fabrication of spintronic devices center on the generation of
stationary spin-polarized currents in non-magnetic semiconductors.

Some of the methods for the generation of stationary spin
polarization are based on spin injection through the interface
between a ferromagnetic metal to a non-magnetic conducting
material; we will refer to this idea as the ``spin-filter'' scheme
\cite{ar}. In the diffusive transport regime, the spin-filter
scheme has been shown initially to be associated with a very small
degree of spin polarization (of the order of a few percent
\cite{ham, hu, hu2, mot}). There are two main reasons for this
inefficiency \cite{mjl1,fil}: (i) the spin relaxation time is much
smaller in a ferromagnetic material than in a non-magnetic one,
and (ii) the conductivity of the ferromagnetic metal injector is
much higher than the conductivity of the semiconductors which are
usually used as non-magnetic materials. In effect, the
nonequilibrium electrons that are injected from the ferromagnet,
undergo a Brownian motion. Consequently, prior to reaching the
detector (collector) these electrons return back into the
ferromagnet repeatedly (or they undergo a spin-flip in the
semiconductor). Because of the high frequency of spin-flip
processes the probability to lose the spin is high in the magnetic
material. Furthermore, due to the aforementioned conductivity
mismatch between the ferromagnetic and nonmagnetic materials, the
electrons will spend most of the time in the ferromagnetic
material, and this will increase the probability to lose the
excess spin orientation. Consequently, the spin polarization of
the current in the semiconductor is expected to be extremely low.

There are number of additional essential limitations inherent to
the spin-filter scheme. First, the spin polarization of the
injected current can not exceed the spin polarization of the
current in the magnetic material (serving as an injector).
Secondly, the distance over which a significant degree of spin
polarization may be maintained in a non-magnetic material, can not
exceed the diffusion spin-flip length in it. In addition, we note
that it is practically impossible to vary the spin polarization of
the injected current, and additional methods are required in order
to detect and /or measure the degree of spin polarization (such as
the use of a light emitting diode \cite{Jon} or the oblique Hanle
effect technique \cite{mot}).

Recently, the spin-injection efficiency has been markedly
increased \cite{mol2,mol3}; indeed, by replacing the ferromagnetic
metal by a diluted magnetic semiconductor (DMS),
Be$_{x}$Mn$_{y}$Zn$_{1-x-y}$Se, a record degree of polarization
(up to 90\%) has been achieved \cite{mol2}. This remarkable result
originates from specific properties of the DMS. In particular,
because of the very large split of the spin subbands in a magnetic
field, these compounds may have a sufficiently high degree of spin
polarization. Consequently, if the Fermi level in the DMS appears
below the bottom of one of the spin subbands, the
spin-polarization may reach 100\%. However, the use of a DMS
instead of a ferromagnetic metal, as well as a number of other
ways suggested recently \cite{mot,han, not}, overcomes only one of
the above mentioned limitations, i.e., they address only the
enhancement of the spin-polarization of the injected current.

In this paper we propose a new method for generation and transport
of high spin-polarized currents. We term the proposed method a
``spin-guide'' scheme. The spin-guide is based on a new interface
configuration \cite{gur1} which allows one to alleviate the
aforementioned intrinsic limitations associated with spin-filter
schemes. Under certain conditions a spin-guide may generate an
almost perfect (100\%) spin-polarized current even when a magnet
with a relatively low degree of spin polarization is used.
Moreover, in the spin-guide scheme spin polarization may be
transmitted over large distances that exceeds significantly the
spin-flip length in non-magnetic materials. Finally, spin- guides
allow easy detection and control of the spin polarization and, as
discussed below, they may form the basis for creating fast
spin-polarization switches.

\section{BASIC IDEA AND APPROACH}

\begin{figure}
\includegraphics[height=5cm,width=8cm]{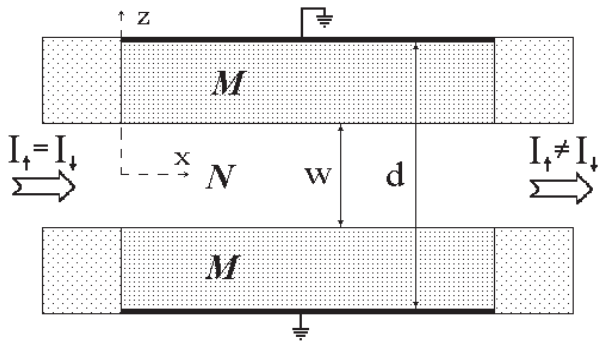}
\caption{Cross-sectional view of the spin-guide scheme. w is the
width of the non-magnetic channel (N), and d is the distance
between the grounded magnetic (M) contacts.} \label{fig1}
\end{figure}

A "spin-guide" is a system consisting of a non-magnetic conducting
channel (wire or strip) wrapped around by a grounded magnetic
shell (see Fig.~\ref{fig1}). Unlike the spin-filter, electric
current flows here along the interface, instead of being normal to
it. The main idea is that nonequilibrium electrons with a
particular spin polarization (e.g., which coincides with the
magnetization axis) leave preferentially the non-magnetic channel
to the magnetic material. The return of these electrons into the
channel is prohibited because the outside magnetic shell
boundaries are grounded. Consequently, a permanent outflow of
nonequilibrium electrons with a definite spin polarization is
obtained, and an excess of nonequilibrium electrons of the other
spin polarization appears in the channel. Note, that the spin
polarization of the current in the channel is opposite to the spin
polarization of the current flowing in the surrounding magnetic
shell, in contrast to the spin-filter geometry.

For the sake of specificity, let us consider a flat configuration
where the interface is a planar plate; the extension to the
three-dimensional case (cylindrical wire) is straightforward. We
will consider the diffusive transport regime, where the diffusion
length $l_{\uparrow, \downarrow }$ (where $l_{\uparrow, \downarrow
}$ are the electron-impurity mean-free paths for the spin-up and
spin-down electrons, correspondingly) are significantly shorter
than any characteristic length of the spin-guide. In this paper
the effects of electron-electron collisions are neglected - this
is \textit{a fortiori} valid at sufficiently low temperatures
(i.e., several degrees Kelvin).

Let $\mu _{\uparrow, \downarrow }$ denote the electrochemical
potentials for the spin-up and spin-down electrons, respectively.
The electric current densities $J_{\uparrow , \downarrow }$ are
related to the electrochemical potentials via Ohm's law

\begin{equation}
\label{eq1}
J_{ \uparrow , \downarrow } = - \frac{{\sigma _{ \uparrow , \downarrow }
}}{{e}}\nabla \mu _{ \uparrow , \downarrow } ,
\end{equation}

\noindent where $\sigma_{\uparrow, \downarrow }$ are the
corresponding conductivities. The spin transport, within the
diffusive regime approximation, is described by Mott's equations:

\begin{equation}
\label{eq2}
\begin{array}{l}
 div(\sigma _{ \uparrow , \downarrow } \nabla \mu _{ \uparrow , \downarrow }
) = \frac{{\Pi _{0} e^{2}}}{{\tau _{sf} }}(\mu _{ \uparrow , \downarrow } -
\mu _{ \downarrow , \uparrow } ), \\
\vphantom{i}\\
 \Pi _{0} ^{ - 1} = \Pi _{ \downarrow } ^{ - 1} + \Pi _{ \uparrow }^{ -
 1}.
 \\
 \end{array}
\end{equation}

Here $\Pi_{\uparrow, \downarrow}$ are the densities of states at
the Fermi level of the up and down spins, and $\tau _{sf}$ is the
spin-flip scattering time. Eqs.(\ref{eq2}) hold under the
assumption that the spin-flip mean free-paths $l_{sf \uparrow,
\downarrow }$= $v_{F \uparrow , \downarrow} \tau_{sf}$ (where
$v_{F \uparrow , \downarrow }$ are the Fermi velocities of the
spin-up and spin-down electrons) exceed significantly the
diffusion lengths $l_{\uparrow, \downarrow }$, i.e.
$l^{sf}_{\uparrow, \downarrow } >> l_{\uparrow, \downarrow }$;
otherwise, the problem should be studied within the kinetic
equation approach. A typical lengthscale on which the equilibrium
between the spin subsystems is established is the diffusive length
$\lambda = (\sigma_0 \tau_{sf}/ e^{2}{\Pi_0})^{1/2}$, where
$\sigma_0^{-1}$ = $\sigma_{\uparrow}^{-1} +
\sigma_{\downarrow}^{-1}$.

Note that we can find the currents in the spin-guide without
separation of the electrochemical potential into the chemical
($\eta $) and electrical ($\varphi $) potential contributions,
i.e., $\mu_{\uparrow, \downarrow }= \eta_{\uparrow, \downarrow } +
e \varphi$. These potentials can be easy obtained from the
solution for $\mu_{\uparrow, \downarrow }$ when the screening
radius is much shorter than the size of the spin-guide (which is
the case in reality). Then, from the condition of electric
neutrality, $\Pi_{\uparrow} \eta_{\uparrow} + \Pi_{\downarrow}
\eta_{\downarrow} = 0$, we have

\[\eta_{\uparrow, \downarrow }= \Pi_{\downarrow, \uparrow}
(\mu_{\uparrow, \downarrow} -\mu_{\downarrow, \uparrow })/\Pi,
\]
\[
 e \varphi =(\Pi_{\uparrow}\mu_{\uparrow} + \Pi_{\downarrow}
 \mu_{\downarrow})/\Pi,
\]
\[\Pi =\Pi_{\uparrow} + \Pi_{\downarrow}.\]

The above equations should be supplemented by the imposed boundary
conditions. Let the $x$-axis be directed along the channel and lie
in it's middle, and take the $z$-axis to be perpendicular to the
interfacial planes, with the origin of the coordinate system
located in the center of the entrance into the channel (see
Fig.~\ref{fig1}). The grounding of the outside boundaries is
equivalent to the condition $\mu_{\uparrow, \downarrow }= 0$.
Taking into account the condition of electric neutrality, we
obtain $\eta_{\uparrow, \downarrow}= \varphi = 0$. It would appear
reasonable to take the same potentials at the channel exit. (An
excess of exit potential over the grounded boundaries is
equivalent to an inefficient removal of energy into the ground.)

Let an unpolarized current I be driven through the channel
entrance. We find that the spin-up and spin-down current densities
may be expressed as $J_{\uparrow, \downarrow}= - e^{-1} \sigma_N
\partial \mu_{\uparrow , \downarrow }/\partial x = I/2w$. As we
show below, at the level of accuracy to which we restrict
ourselves in this paper, the result is insensitive to the type of
boundary conditions that are imposed at the end faces of the
magnetic shell (grounding, or absence of current). We also assume
that the conductivity in the non-magnetic channel $\sigma _{N}$ is
spin independent, and that the conductivity in each region is
constant.

For the spin-guide model described above, the diffusion equation
can be solved exactly. Instead of quoting here the general
solution (which is rather complicated) it is sufficient for our
purpose to focus attention on the solutions that hold far way from
the ends of the spin-guide. We also assume that the length of the
non-magnetic channel, L, is much larger then the width of the
spin-guide, i.e. $L \gg d$, and solve Eqs.(\ref{eq2}) through a
separation of variables, by introducing the new functions $\mu_{+}
= (\sigma_{\uparrow} \mu_{\uparrow} + \sigma_{\downarrow}
\mu_{\downarrow})/(\sigma_{\uparrow}+ \sigma_{\downarrow})$ and
$\mu_{-} =\mu_{\uparrow}  -  \mu_{\downarrow} $.

Due to the symmetry of the system and the boundary conditions at
$z= \pm d/2$, we obtain the following solution

\begin{equation}
\label{eq3}
 \mu_{\pm} =e^{-kx}f_{\pm}(z),
\end{equation}
where the functions f$ \pm $ are given by:

\begin{equation}
\label{eq4}
\begin{array}{l}
 f_{ - } = \left\{
 \begin{array}{ll}
 C \ cos(\kappa_{N} z),          &  |z|< w/2, \\
 D \ sin(\kappa_{M}(d/2 - z)),   &  |z| >  w/2,\\
 \end{array} \right.  \\
 \\
 f_{ + } = \left\{
 \begin{array}{ll}
 A \ cos(kz),          &         |z| < w/2, \\
 B \ sin(k(d/2 - z)),  &         |z| > w/2, \\
\end{array}  \right. \\
 \\
 \end{array}
\end{equation}
and
\[
\kappa _{M,N} = \sqrt {k^{2} - \lambda _{M,N}^{ - 2} } \quad ,
\]
where $\lambda_{M,N }$ is the diffusion length in the magnetic (M)
and non-magnetic (N) regions, respectively. Matching the functions
$\mu_{\uparrow , \downarrow}$and the current (i.e. the derivatives
$\sigma_{\uparrow, \downarrow} \partial \mu_{\uparrow,
\downarrow}/\partial z$) at $z =  \pm w/2$, we can determine the
values of the coefficients A,B,C and D (in terms of the value of
one of them), as well as find the possible values of the damping
factor k. To exponential accuracy, it is sufficient to consider
only the solution with the smallest value of $k$. The physical
meaning of $k^{-1}$ is quite obvious: it is the distance in the
$x$-direction which an electron will traverse diffusively before
it will reach the grounded contact.

In general the solutions given by Eqs.(\ref{eq3}) are not
applicable at distances from the channel ends which are less than
$k^{-1}$. But, under the assumption that $L >> k^{-1}$, and to the
accuracy of our analysis, we may use these solutions even near the
channel exit. As shown below this amounts to a neglect of a
preexponential factor in the expression for the current spin
polarization.

Before closing this section, let us consider another type of
solution which is valid for distances from the entrance where
spin-flip processes have not yet occurred. In the absence of
spin-flip Eqs.(\ref{eq2}) for $\mu_{\uparrow} $ and
$\mu_\downarrow$ become independent and a separation of variables
can be accomplished separately for each potential. Thus, we have
\begin{subequations}
\label{eq5}
\begin{flalign}
 \mu_{\uparrow,\downarrow}=e^{-k_{\uparrow, \downarrow}x}
 f_{\uparrow,\downarrow}, & &\\
 \intertext{with}
 f_{\uparrow, \downarrow}= A_{\uparrow, \downarrow}
 cos(k_{\uparrow, \downarrow} z)& \qquad \mbox{at}\qquad |z|< w/2 ,\nonumber\\
 \vphantom{a}& &\\
 f_{\uparrow, \downarrow }=B_{\uparrow, \downarrow}
 sin(k_{\uparrow, \downarrow } (d/2-z)) & \qquad \mbox{at} \qquad |z|>w/2.  \nonumber
 \end{flalign}
\end{subequations}

From the matching conditions at $z =  \pm w/2$ the following
transcendental equations are obtained for the damping factors

\begin{equation}
\label{eq6}
 tan(k_{\uparrow, \downarrow } w/2)tan(k_{\uparrow, \downarrow}
 (d-w)/2))= \sigma_{M \uparrow , \downarrow }/ \sigma _{N}.
\end{equation}

\bigskip

In the next section we use the above solutions Eqs. (\ref{eq3})
-(\ref{eq6}) in the analysis of several limiting situations for
different spin-guide parameters.

\section{RESULTS}

\subsection{A fully polarized magnetic region}

A most effective implementation of the spin-guide involves the use
of a DMS with a very large Zeeman splitting as the magnetic
environment, so that the electrons in the magnetic material are
fully spin-polarized. Clearly, spin-flip process in the magnetic
region are precluded in this case. For definiteness, let us assume
that the magnetic shell is not transparent for "spin-up"
electrons, i.e. $\sigma _{M \uparrow } = 0$.

We consider the case when the spin polarization of the current in
the channel is high enough, i.e. the width of the non-magnetic
channel w is less than the spin-flip length $\lambda _{N}$. This
situation is quite real; in particular, we note that since the
spin-flip process is of relativistic origins it is characterized
by a large spin-flip length in non-magnetic semiconductors, i.e.
up to 100 $\mu $m \cite{hag, kik}. For distances from the entrance
short enough so that no spin-flip processes have occurred, the
current $I_\uparrow $ will be conserved inside the channel (that
is, it does not depend on $x$). On the other hand, the current of
electrons with the opposite spin direction, $I_\downarrow$ will
decrease exponentially with distance from the entrance into the
channel, i.e. $I_{\uparrow}  \propto exp(-k_{\downarrow} x)$.

According to Eq.(\ref{eq6}) we have $k_\uparrow  = 0$, and $k_
\downarrow $ will depend on the ratio $\sigma _{M \downarrow }/
\sigma _{N}$. Accordingly, for $\sigma_{M \downarrow }=
\sigma_{N}$ the damping factor $k_\downarrow  = \pi /d$. If the
conductivity of the magnetic shell is much higher than that of the
non-magnetic channel, i.e. when $\sigma _{M \downarrow }>> \sigma
_{N}$, the damping factor takes the value $k_\downarrow  =
\mbox{min} \{ \pi /w, \pi /(d - w) \}$. Consequently, the spin
polarization of the current at the channel exit tends
exponentially to unity with increasing $L$, that is
\begin{equation}
\label{eq7}
\alpha = \frac{{I_{ \uparrow } - I_{ \downarrow }
}}{{I_{ \uparrow } + I_{ \downarrow } }} \approx 1 - e^{ - k_{
\downarrow } L}.
\end{equation}

Note that the difference between $\alpha $ and unity, which decreases for
larger distances, is determined here only within a pre-exponential factor.

We turn now to analysis of the role of spin-flip processes in the
non-magnetic channel. Using Eqs.(\ref{eq3}) and (\ref{eq4}) we
obtain
\begin{equation}
\label{eq8}
k^{-1}=\lambda _{N}
\end{equation}
and
\begin{equation}
\label{eq9} 1 - \alpha =\frac{w(3d-2w)}{12\lambda_N^2}
\le\frac{wd}{\lambda_N^2} \ll 1.
\end{equation}

Thus, the exponential decrease of $1 - \alpha $ (recall
Eq.(\ref{eq7})) is bounded below by the value given in
Eq.(\ref{eq9}). Consequently, the spin polarization remains
constant and sufficiently high for all distances away from the
entrance. However, both the spin-up and spin-down currents
$J_{\uparrow, \downarrow }$ will decay exponentially with the same
damping factor $k$. The total current will decay as the spin-up
electrons succeed in leaving the non-magnetic channel due to
spin-flip processes.

\subsection{A non-ideal magnetic region}

In this section, we discuss the situation when the magnetic shell which
surrounds the conducting non-magnetic channel is not fully polarized - in
this case both spin-up and spin-down currents flow through the shell and
spin-flip processes are possible. The coefficient of selective transparency
of the magnetic shell is determined by the relation

\begin{equation}
\label{eq10}
\gamma = \frac{{\sigma _{M \uparrow } }}{{\sigma _{M
\downarrow } }} < 1.
\end{equation}

This parameter determines the upper bound value of the spin
polarization $\alpha = (1-\gamma )/(1+\gamma )$ in the spin-filter
scheme. For simplicity, we will neglect in the following spin-flip
processes in the non-magnetic channel.

We consider first the case where we may neglect the spin-flip
processes in the magnetic shell near the entrance to the
spin-guide. Then, according to Eq.(\ref{eq5}), the spin
polarization of the current in the channel will tend exponentially
to the unity,
\begin{equation}
\label{eq11} \alpha \approx  1 - e^{-(k_ \downarrow -k_\uparrow
)x}.
\end{equation}

Moreover, from Eqs. (\ref{eq10}) and (\ref{eq6}) we have
$k_\downarrow > k_\uparrow $. As shown in the previous section,
$k_\downarrow^{-1} \leqslant  d$, and for $\sigma _{M \uparrow }~
\ll~ \sigma _{N}$ the spin-up current decays on a length-scale
which is large compare to $d$, i.e.

\begin{equation}
\label{eq12}
k_{ \uparrow } = 2\sqrt {\frac{{\sigma _{M \uparrow }
}}{{\sigma _{N} w(d - w)}}} \quad .
\end{equation}

Now we consider the role of spin-flip processes in the magnetic
shell. As discussed above, the exponential decrease of the
currents $J_{\uparrow, \downarrow}$ (as a function of distance
away from the channel entrance) occurring with the corresponding
damping factors $k_{\uparrow, \downarrow}$, will be changed due to
the spin-flip processes in such a way that both the up and down
components of the current will decrease with the same damping
factor k. Assuming that the diffusion of the electrons to the
grounded boundaries occurs with a faster rate than the spin-flip
processes, i.e. that the condition $\lambda _{M} \gg (d-w)$ is
fulfilled, we obtain (to a first approximation) that the damping
factor k is the same as k$ \uparrow $ determined from the
Eq.(\ref{eq6}). The reason is that the overall damping rate is
governed by that component which takes more time to reach the
grounded boundaries. The spin polarization $\alpha $ which arises
at such length-scale (measured from the channel entrance) can be
found by matching the solutions of Eq.(\ref{eq4}) and expanding
them in terms of the small parameter. Thus, we obtain:

\begin{equation}
\label{eq13} 1 - \alpha = \frac{{\gamma }}{{2(1 - \gamma
^{2})(k\lambda _{M} )^{2}}} \cdot \left( {\frac{{k(d -
w)}}{{sink(d - w)}}} \right. - 1\left. {{\begin{array}{*{20}c}
 {} \hfill \\
 {} \hfill \\
\end{array} }} \right).
\end{equation}

In conjunction with Eq.(\ref{eq6}) for $k=k_\uparrow $, Eq.
(\ref{eq13}) determines a high enough degree of the spin
polarization:
\[
1 - \alpha  \quad  \approx  \quad \gamma  (d-w)^{2}/\lambda
^{2}_{M}(1 - \gamma ^{2}) << 1.
\]

We remark that this inequality will be violated if $\gamma $ is too close to
the unity, i.e. in this case our expansion is inapplicable.

In conclusion, we obtained that in the spin-guide scheme the spin
polarization of the current may be propagated over arbitrarily
long distances, in contrast to the spin-filter scheme where the
transport length-scale is of the order of the diffusion spin-flip
length $\lambda $. There are additional essential differences
between the two schemes. Unlike the spin-filter scheme, the spin
polarization $\alpha $ in the spin-guide does not depend on the
ratio $\sigma _{M \uparrow } / \sigma _{N}$. Moreover, as may be
seen from Eqs.(\ref{eq11}) and (\ref{eq13}) the degree of spin
polarization in the non-magnetic channel can exceed significantly
the degree of spin polarization in the magnetic material.

In the case that $\lambda _{M} \gg d$, a sufficiently high degree
of spin polarization may be achieved when the condition $\gamma
(d-w) << \lambda _{M }$ is fulfilled, i.e.

\begin{equation}
\label{eq14} 1 - \alpha = \gamma \frac{{tan \ k \ (d -
w)}}{{2k\lambda _{M} }} \approx \gamma \frac{{d - w}}{{\lambda
_{M} }} < < 1.
\end{equation}

If the magnetic shell is too thick, i.e. when $\gamma (d-w)
/\lambda _{M} \gg 1$, then the spin polarization of the current
will be low.

As aforementioned, to increase the spin polarization one should
decrease the width of magnetic region. To this end, the ballistic
regime when $(d-w)  \ll l_{M \uparrow },\quad l^{sf}_{\uparrow,
\downarrow }$ is most favorable. A calculation which goes beyond
the framework of the diffusion approach yields in the ballistic
limit the following result:

\begin{equation}
\label{eq15} 1 - \alpha \approx \gamma \frac{{l_{M \downarrow } (d
- w)}}{{\lambda_{M}^2 }}\quad ln\frac{{l_{M \downarrow }}}{{(d -
w)(l_{M \downarrow }^2 \lambda _{M} ^{ - 2} + 1)}}.
\end{equation}

The logarithmic factor in this formula reflects an enhancement of the
spin-flip probability for electrons grazing along a magnetic layer.

\section{SPECIFIC EFFECTS AND POSSIBLE EXPERIMENTAL
REALIZATIONS}

In this section we consider some possible experimental schemes aiming at
realization of the proposed transport phenomena, and at direct observation
of the spin polarization of the current flowing in a spin-guide.

\subsection{Spin drag}

\begin{figure}
\includegraphics[height=6 cm,width=8cm]{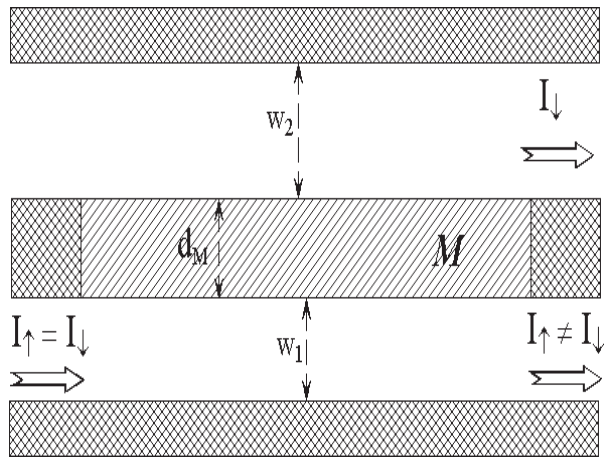}
\caption{The spin-drag scheme. The semiconducting non-magnetic (N)
channels 1 and 2 are of widths $w_{1}$ and $w_{2}$, respectively.
respectively. The width of the magnetic (M) interlayer is
$d_{M}$.} \label{fig2}
\end{figure}

We begin with a discussion of an alternative scheme to the one
discussed above, for obtaining the spin-guide effect on the
polarization of the electric current. This alternative scheme is
based on a new spin-drag effect that can be realized in a geometry
of two semiconductor channels separated by a magnetic interlayer
(see Fig.~\ref{fig2}).

Let a non-polarized current enter into the nonmagnetic
(semiconductor) channel 1. In the case of a fully polarized
magnetic interlayer a fully polarized current will appear in the
semiconductor channel 2 due to the spin-filter effect, i.e.
$|\alpha _{2}| = 1$. At the same time, a polarization $\alpha
_{1}$ will appear in the first channel due to the spin-guide
effect, and it's magnitude will depend on the relative width of
the channels. If the thickness of the magnetic interlayer is taken
to be less than the values of $w_{1}$, $w_{2 }$ and $ L \gg
w_{1}$, $w_{2 }$ (where $L$ is the channel length), we have

\begin{equation}
\label{eq16}
\alpha _{1} = \frac{{w_{2} }}{{w_{2} + 2w_{1}^{} }}.
\end{equation}

The polarizations of the currents in channels 1 and 2 are opposite
to each other, and the total current in the two channels is
non-polarized. It is of interest to note that if channel 2 is
sufficiently wide such that $w_{1} \ll w_{2}$, the polarized
currents will be equally divided between the channels, i.e. an
entirely polarized current $J_\uparrow  = J_{0}/2$ will appear in
channel 1, with an equal value and opposite polarization to that
in channel 2. In the derivation of Eq.(\ref{eq16}) we have assumed
that the potential applied at the exit of channel 2 is the same as
that applied at the exit of channel 1 (the latter is determined in
our model by the value of the current $J_{0}$). Varying the
potential at the exit of channel 2, one can control the current
polarization in the channels.

In case that the magnetic interlayer is not fully polarized, but
$\gamma  \ll 1$, the spin polarization determined by
Eq.(\ref{eq16}) is conserved at distances $L~<~R$, where
$R~=~\mbox{min}\{(rw\sigma _{N}/ \sigma _{M \uparrow })^{1/2}$,
$\lambda _{N}\}$ and $r~=~\mbox{min} \{ d_{M}, \lambda _{M}, l_{M
\uparrow }\}$. Here we take into account the possibility that the
propagation of the electrons in the magnetic interlayer is either
diffusive or ballistic. If the magnetic interlayer is wider than
the non-magnetic channels, i.e. $d_{M} \quad _{ }> w_{1}$,
$w_{2}$, then the Sharvin resistances of the exit constrictions of
the system should be used in the expressions for $\alpha _{1,2}$.

The above considerations lead us to suggest the creation of a fast switch of
the spin polarized current, achieved by combining the spin drag scheme with
electrostatic gates at the exits of the channels one may switch the spin
polarization of the current at a fast rate without switching the
magnetization of the magnetic material. In the spin-filter scheme fast
switching of the current spin polarization is unachievable even under the
best conditions, i.e. when using DMS structures. This is because of the
required applied high magnetic fields, and the comparatively long relaxation
times of the atomic magnetic moments. On the other hand, as mentioned above
the direction of the spin polarization in the spin-guide is opposite to that
appearing in the spin - filter scheme at the same polarization of the
magnetic material. Therefore, it may be possible to create an alternative
switching scheme that combines the spin-filter and the spin- guide schemes
with controllable electrostatic gates.

\subsection{Giant magnetoresistance and direct measurement of
the current spin polarization}

In this section we discuss certain physical effects which could be utilized
for the detection and measurement of the current spin polarization.

A spin-guide consisting of a DMS magnetic shell should exhibit a
giant magnetoresistance effect. The effect is associated with the
decrease of the conductance in the channel (to an essentially
vanishing value) upon switching-off of the magnetizing field; the
reverse happens when the magnetizing field is switched-on, i.e. a
current appears in the channel. This is caused by the fact that
the disappearance of all the nonequilibrium electrons at the
grounded boundaries is faster then the rate of their arrival to
the channel exit. This effect may result in a most significant
change of the resistance of the device with a magnetizing field,
perhaps even larger then the the giant magnetoresistance effect
measured in the spin-filter scheme \cite{mol3,bab}.

If the ferromagnetic material which surrounds the non-magnetic
channel in a spin- guide is not fully polarized, a giant
magnetoresistance effect may be observed for the case of mutually
opposite magnetizations of the upper and lower magnetic layers
(see Fig.~\ref{fig1}) ) to avoid the residual magnetization. If
the upper and lower magnetic layers have the same magnetization
then there is a current at the channel exit, but if their
magnetizations are opposite then the current will essentially
vanish. Therefore, by changing the applied magnetic field we may
change the resistance of the device.

\begin{figure}
\includegraphics[height=6 cm,width=8cm]{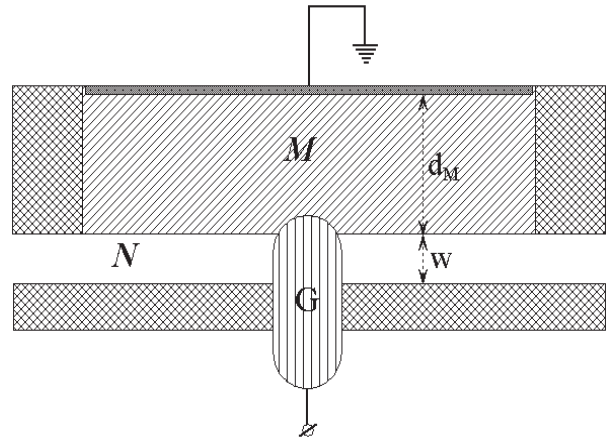}
\caption{Scheme of the experiment with a locking of the
nonmagnetic channel in a spin guide. G is an electrostatic locking
gate, $d_{M}$ and w are the widths of the magnetic (M) and
non-magnetic (N) materials, the double hatching indicates the
dielectric regions.} \label{fig3}
\end{figure}

Another spin guide effect with may be observed by locking a
non-magnetic channel far from the entry and exit by an
electrostatic gate, as shown schematically in Fig.~\ref{fig3}. In
this case, the essential variation of the current indicates on the
effectiveness of the spin guide.

At last, in spin-guide with blocked channel and fully polarized
magnetic shell one can measure the spin-polarization $\alpha $
directly:

\[
\alpha \; = 1 - \frac{{I\left( {B \ne 0} \right)}}{{I\left( {B = 0}
\right)}}
\]
here ${I(B \ne 0)}$, $I(B=0)$ are currents at exit of spin-guide
at switched-on and switched-off magnetized magnetic field
correspondingly, both in case of locked gate.

The spin polarization of the current in a spin-guide with a DMS
shell can be measured directly by locking a non-magnetic channel
far from the entry and exit by an electrostatic gate, as shown
schematically in Fig.~\ref{fig3}. In this case the relative change
of the current $\delta I/ I$ at locking of the channel is in
direct proportion to the spin polarization, $\alpha $, and to the
ratio $w /d_{M}$, where $d_{M}$ is the width of a magnetic layer
(M).

\section{DISCUSSION}

The main operational principle of a spin-guide is the removal of
one of the components of the spin current from the channel due to
the selective transparency (with respect to the spin direction) of
a magnetic shell. The spin polarization of the current increases
with distance from the channel entrance until spin-flip processes
become effective. Thus, in contrast to the spin-filter scheme, the
spin polarization in a spin-guide can exceed significantly the
spin polarization of the current in the magnetic material which
surrounds the non-magnetic channel. In general, a spin-guide may
generate an almost fully (100\%) spin-polarized current even if
the magnetic material which is used is not fully polarized. Even a
small difference between the spin-up and spin-down conductivities
in the magnetic material ($\sigma _{M \uparrow }/\sigma_{M
\downarrow } < 1$ in our case) would lead to a depletion of the
current states in the non-magnetic channel, with spin-down
electrons being affected over shorter distances from the channel
entrance than the spin-up electrons. In this case, the spin
polarization will be determined by the difference of the
quantities in the exponent of Eq.(~\ref{eq11}); consequently, the
spin polarization of the current will tend to approach the
limiting value (i.e. 100\%) further into the channel.

At this stage, certain issues pertaining to the operation of the
proposed spin-guide scheme warrant comment. We begin by noting
that though the spin polarization is expected to remain high even
when a non-ideal magnetic material is used, the total current in
the channel will decrease with increasing channel length (see
Eqs.(\ref{eq1}), (~\ref{eq3}) and (~\ref{eq5})). This occurs
because both the spin-up and spin-down electrons can leave the
channel and thus reach the grounded contact. In this context we
note that there are ways to reduce significantly the loss of
current, thus allowing its at transmission over a large distance.
To this end one may wish to use an alternation of the grounded and
ungrounded sections of the magnetic shell along the spin-guide.
Alternatively, one may reduce the loss of total current in the
channel by creating tunnel barriers between the non-magnetic
channel and the magnetic shell; such barriers, however, will
retard the exit of both electron polarizations to the grounded
boundaries.

The polarizing ability of a spin-guide is limited only by the
spin-flip processes. Here we should note, that the role of
spin-flip processes both in a non-magnetic channel and in the
magnetic region of the spin-guide, differs in an essential way
from the role of spin-flip processes in a spin-filter. First, let
us consider spin-flip processes in the non-magnetic semiconductor
only. In contrast to the spin-filter scheme, while spin-flip
limits the spin-polarization in the spin-guide, it can not destroy
it fully. Moreover, the spin polarization remains a constant and
high enough, as follows from Eq.(~\ref{eq9}), for arbitrarily
large distance from the entrance.

Next we consider the role of spin-flip in the magnetic shell of a
spin-guide. Generally speaking, the exit of electrons with a spin
``parallel'' to that in the magnetic region (spin-down in our
case), from the non-magnetic channel into the magnetic
surroundings (as a result of their Brownian motion) is a harmless
useful process. It is obvious that spin-flip scattering of these
electrons will not reduce the spin polarization in the
non-magnetic channel. However, spin polarization will be reduced
due to the exit from the channel of spin-up electrons. They could
change the spin polarization due to spin-flip scattering in the
magnetic region and, subsequently, they could return back in the
non-magnetic channel. The spin-flip probability could be decreased
by reducing the width of the magnetic region; this method of
bringing about a decrease in the spin-flip probability is not
possible for the spin-filter scheme. In fact if the magnetic shell
width is less than $\lambda_{M}$ the sources of nonequilibrium
spin concentration at the entrance and the exit of the current in
the magnetic region will mutually cancel each other, and the
polarizing ability of the magnetic filter will decrease
significantly, as observed experimentally \cite{mol2,mol3}.
Furthermore, the high conductivity of the magnetic material in the
spin-guide scheme does not increase the spin-flip probability
because it speeds up the transport of electrons to the ground
contact. Unlike the spin-filter scheme, spin polarization in the
spin-guide, $\alpha $, does not depend on the ratio $\sigma _{M
\uparrow } / \sigma _{N \uparrow }$. We recall that the large
ratio $\sigma_{M \uparrow } / \sigma_{N \uparrow }$ ,
characteristic of the spin-filter ``ferromagnetic
metal-semiconductor'' interface \cite{ham, hu, hu2, mot}, is one
of the main reasons for the low degree of spin polarization in
this scheme \cite{mjl1}. If the spin-guide is used with tunnel
barriers between the non-magnetic channel and the magnetic shell
and with an additional applied voltage to the tunnel barrier, then
the barriers act as additional filters. Those electrons that
crossed the barriers and underwent an inelastic scattering in the
magnetic shell are not capable to return back into the
non-magnetic channel. Consequently, the spin-flip processes in the
magnetic region will affect the current spin polarization in the
channel to a lesser extent.

Thus, there is a physical difference in the role of spin-flip between the
two schemes. Spin-flip scattering in the magnetic shell of spin-guide leads
mainly to a reduction of the total current, while the spin polarization may
change only by a small amount. The reverse situation occurs in the
spin-filter scheme, i.e. the spin-flip processes maintain the total current
as a constant but cause a significant reduction in the spin-polarization, as
discussed in Section I.

As evident from the above, the spin polarization of the current in a
spin-guide depends significantly both on it's length and on the widths of
the channel and the magnetic shell. Hence, by varying these parameters it
should be possible to readily change and control the degree of current
polarization at the channel exit. In the following we provide some
quantitative estimates concerning the degree of spin polarization that may
be achieved in the spin-guide scheme.

Among the most promising candidates for the magnetic shell
material in a spin-guide are II-VI-DMSc compounds (like a
$Be_{x}Mn_{y}Zn_{1-x-y}Se$) or halfmetals where one of the spin
subbands can be fully pinned. Assuming a non-magnetic channel with
$\lambda _{N} = 1.5 \mu$m (a case that is far from being optimal),
$=0.3 \mu$m and $d=0.4 \mu$m, we obtain according to the
Eq.(\ref{eq9}) a current spin polarization in the channel $\alpha
 = 100\%$ (within a $1\%$ accuracy) for an arbitrary distance from
the entrance; the current amplitude will decay with $\lambda
_{N}$, according to Eq.(\ref{eq9}). Even for DMS compounds which
are not fully polarized, employed as the magnetic region, we can
achieve a high spin polarization. For example, taking
$Zn_{0.97}Be_{0.03}Se$ as a NMS (non-magnetic semiconductor)
channel material, in contact with a $45\%$ polarized
$Zn_{0.89}Be_{0.05}Mn_{0.06}Se$ as a DMS shell with a spin-flip
length $\lambda  \quad  \approx  20$nm, yields according to
Eqs.(\ref{eq13}) and (\ref{eq14}) a $95\%$ spin polarization for a
width of the magnetic shell $(d-w)  \leqslant  10$nm; for $(d-w)
\approx  50$nm we obtain a spin polarization $\alpha \approx
17\%$.

Finally, a very high degree of spin polarization of the current
may be achieved even if a ferromagnetic metal shell (e.g., Ni , Fe
or Py) is used in the spin-guide. Here one should employ thin
ferromagnetic films with a thickness that is less than the
diffusion spin-flip length $\lambda _{M}$; this is feasible even
when $\lambda _{M}$ is about several \textit{tens of }nanometers.
Thus, when the ballistic regime is reached in the magnetic region,
f.e. $\lambda _{M} \quad  \approx  20$nm \cite{dub,ans}, with
$\gamma \approx  0.6$, $d-w  \approx 8$nm and $\lambda _{N} \quad
\approx  1.5 \mu$m, one obtains from Eq.(\ref{eq15}) that $\alpha
\quad \approx 100\%$, within the accuracy of the model. For rather
thick film, such that the diffusion regime is reached, with $d =
60$nm, $w = 0.7 d$, $\lambda _{M} = 20$nm, we obtain from
Eq.(\ref{eq13}) $\alpha  \approx 0.97\%$.

From the above we conclude that the spin-guide scheme works most effectively
if both the widths of the non-magnetic channel and the magnetic shell are
taken to be much less than the corresponding spin-flip length. In view of
realistic spin-flip length scales, we suggest that nano-scale structures
would be most appropriate for fabrication of spin-guide devices; for example
through the use of nanowires and layers of nano-widths dimensions.

\section{SUMMARY}

In this paper a spin-guide has been proposed as a new type of
source and a long-distance transmission medium of electric
currents with a high degree of spin polarization. We have shown
that spin-guide enhances significantly the capabilities for
generation and manipulation of spin-polarized currents. The main
features of the spin-guide scheme which make it a most promising
tool for creation and transport of spin-polarized currents in
non-magnetic semiconductors, may be summarized as follows:

(i) In a spin-guide, a permanent withdrawal of electrons of one spin
polarization leads to an increase in the other spin polarization, thus
allowing to achieve a high degree of spin-polarization of the current,
considerably exceeding the degree polarization in the magnetic shell.

(ii) The propagation length of a spin - polarized current in a non-magnetic
channel of a spin-guide may exceed significantly the spin-flip length in the
material.

(iii) Spin-flip processes in the magnetic shell restrict the peak value of
the spin polarization of the current in a spin-guide to a much lesser degree
than in the spin-filter scheme.

(iv) Through the use of the spin-drag scheme (or by combining the spin-guide
and spin-filter schemes) with electrostatic gates on the channel exits, it
may be possible to control the spin polarization of the current and to
switch it easily and promptly, without magnetization reversal of the
magnetic shell.

(v) A very large magneto-resistance effect is predicted to occur, which
should allow direct sensing and measurement of presence and degree of spin
polarization.

\section{Acknowledgements}

The research described in this publication was made possible in part by
Award No.UP2-2430-KH-02 of the U.S. Civilian Research \& Development
Foundation for the Independent States of the Former Soviet Union (CRDF), and
by the US Department of Energy Grant FG05-86ER-45234.

\end{document}